\documentclass[pre,showpacs,superscriptaddress,floatfix,twocolumn]{revtex4}
\usepackage{amsfonts}
\usepackage{graphicx}
\usepackage{slashbox}

\newcommand{\ep}{\epsilon}

\newcommand{\xv}{\vec{x}}

\newcommand{\pd}[2]{\frac{\partial #1}{\partial #2}}

\newcommand{\be}{\begin{equation}}
\newcommand{\ee}{\end{equation}}
\newcommand{\bea}{\begin{eqnarray}}
\newcommand{\eea}{\end{eqnarray}}

\begin{document}

\title{Constant Flux Relation for diffusion limited cluster--cluster aggregation}
\author{Colm Connaughton}
\email{connaughtonc@gmail.com} 
\affiliation {Centre for Complexity Science, University of Warwick, Gibbet Hill
Road, Coventry CV4 7AL, UK}
\affiliation{Mathematics Institute, University of Warwick, Gibbet Hill
Road, Coventry CV4 7AL, UK}
\author{R. Rajesh}
\email{rrajesh@imsc.res.in}
\affiliation{
Institute of Mathematical Sciences, CIT Campus, Taramani, Chennai-600 113,
India}
\author{Oleg Zaboronski}
\email{O.V.Zaboronski@warwick.ac.uk}
\affiliation{Mathematics Institute, University of Warwick, Gibbet Hill
Road, Coventry CV4 7AL, UK} 

\date{\today}

\begin{abstract}
In a non-equilibrium system, a Constant Flux Relation (CFR) expresses the fact that a constant flux of a conserved quantity exactly determines the scaling of the particular correlation function linked to the flux of that conserved quantity. This is true regardless of whether mean--field theory is applicable or not. We focus on cluster--cluster aggregation and discuss the consequences of mass conservation for the steady state of aggregation models with a monomer source in the diffusion-limited regime. We derive the CFR for the flux-carrying correlation function for binary aggregation with a general scale-invariant kernel and show that this exponent is unique. It is independent of both the dimension and of the details of the spatial transport mechanism, a property which is very atypical in the diffusion-limited regime. We then discuss in detail the ``locality criterion'' which must be satisfied in order for the CFR scaling to be realisable. Locality may be checked explicitly for the mean-field Smoluchowski equation. We show that if it is satisfied at the mean-field level, it remains true over some finite range as one perturbatively decreases the dimension of the system below the critical dimension, $d_c=2$, entering the fluctuation-dominated regime. We turn to numerical simulations to verify locality for a range of systems in one dimension which are, presumably, beyond the perturbative regime.  Finally, we illustrate how the CFR scaling may break down as a result of a violation of locality or as a result of finite size effects and discuss the extent to which the results apply to higher order aggregation processes.
\end{abstract}

\maketitle

\section{\label{sec-intro} Introduction and Motivation}

Consider a collection of particles undergoing some spatial transport process which, upon encountering each other, coalesce irreversiby with some probability. Such a situation arises in a great variety of seemingly unrelated branches of science (see \cite{FRI2000} for an overview). Some of the most obvious examples are found in astrophysics \cite{KON01}, aerosol physics \cite{DRA72} and polymer chemistry \cite{ZIF80}. Less obvious examples arise from granular media \cite{CLMNW96}, the structure of drainage networks \cite{DR99,HUB91} and sandpile models of self-organised criticality \cite{DHA99}. This diverse range of applications is one reason why models of systems of diffusing particles which aggregate upon contact have been extensively studied since the seminal work of Smoluchowski laid the foundations for their analysis. A second reason for the enduring interest shown by the scientific community in aggregation models is that they provide simple examples of a surprising array of non-trivial phenomena in non-equilibrium statistical mechanics making them an attractive theoretical proving ground.

Two situations are commonly encountered, depending on the application. One may start with a specified initial distribution of cluster sizes and study how it decays in time. This is sometimes referred to as free aggregation. Alternatively one may start with an empty system and add monomers at a given rate. This is called aggregation with a source. Due to irreversibility of the coagulation process, free aggregation is an entirely dynamic problem with no stationary state. On the other hand, aggregation with a source may produce a stationary distribution of particle sizes in the limit of large time. Stationarity comes about as follows: the rate of decrease of the density of clusters of a given size via coagulation to form larger ones is balanced by the generation of clusters of that size via coagulation of smaller ones. Such a balance is possible only because the source continually replenishes the available pool of small clusters. Clearly such a stationary state is not an equilibrium state since there is no detailed balance. Rather it is a flux state characterised by a constant flux of mass through the space of cluster sizes. On a technical note, since both diffusion and aggregation conserve total mass, the constant influx of monomers results in a linear
increase in the average mass.
While this driving occurs at the smallest mass in the problem, the
aggregation process transfers this mass to larger and larger mass scales. Thus, strictly speaking, such systems are quasi-stationary at large times: small masses reach a stationary distribution but time-evolution proceeds indefinitely at the largest masses.  To attain a truly stationary state, one should introduce a cut-off at some large cluster size above which clusters are removed from the system. In this paper, we concern ourselves exclusively with aggregation problems with a source.

The most basic quantity of interest is the average mass density, $\langle N(m,x,t)\rangle$, which tells us the average number of clusters of a given mass, $m$. Typically, a system of aggregating particles exhibits two regimes of behaviour as a function of the spatial dimension. A critical dimension, normally two for systems undergoing diffusive transport, separates these regimes. In higher dimensions, the dynamics is typically reaction limited and a mean-field description is appropriate. This mean-field description is given by the Smoluchowski kinetic equation which describes the time evolution of $N(m,x,t)$. In lower dimensions, the dynamics is typically diffusion limited. Diffusive fluctuations are strong and a mean-field description is no longer possible. A huge amount is known about the average mass density in the mean-field case \cite{ALD1999} from exact analyses \cite{LEY2003} and extensive numerical simulations of the Smoluchowski equation. Relatively less is known about the mass density in the diffusion limited regime but several models have been solved exactly or treated approximately by field-theoretic methods \cite{CRZ2005,CRZ2006}. Almost nothing is known about higher order correlation functions in the diffusion-limited regime, despite the fact that they encode the details of the fluctuations which dominate the dynamics. This paper concerns itself with such higher order correlation functions, albeit some rather special ones.

The special correlation functions which we consider, can be referred to as the flux-carrying correlation functions. For a given aggregation model with source which attains a constant-flux stationary state at large times, there is a particular correlation function associated with mass transfer. In the turbulence literature, where transfer of energy is analogous to transfer of mass, it is very well known that constancy of the energy flux determines exactly the scaling of the flux-carrying correlation function (see chap. 6 of \cite{frischBook} for example). This fact is the basis for the Kolmogorov $4/5$-Law for three-dimensional hydrodynamic turbulence and its various incarnations in other turbulent systems.  While the $4/5$-Law has become central to the modern understanding of turbulence,  the fact that a similar exact result is available for other non-equilibrium systems, in particular for aggregation systems, has hardly been taken advantage of. The purpose of the present article is to address this issue. In previous work \cite{CRZ2007}, we showed how a conservation law leads to an exact scaling exponent for the flux-carrying correlation function for a broad class of non-equilibrium systems which included aggregation, referring to such a constraint as a ``Constant Flux Relation'' (CFR). In the present article we focus entirely on the consequences of CFR for aggregating particle systems, leaving the original hydrodynamic analogy behind. In the process of verifying the CFR for a broad set of aggregation models in the diffusion-limited regimes we will present a number of somewhat counter-intuitive numerical results which would be very difficult to understand without any prior understanding of the CFR.

The layout of the paper is as follows. We first define the model and give a heuristic derivation of the CFR scaling (Sec.~\ref{sec-model}). We then provide an accurate derivation (Sec.~\ref{sec-cfr}) which makes explicit the assumptions involved, in particular the assumption of locality which we then discuss in detail (Sec.~\ref{sec-locality}). Sec.~\ref{sec-simulations} then reports the results of a large number of numerical simulations which verify the CFR scaling for a range of aggregation kernels, expose finite size effects, test the locality condition in the diffusion limited regime in one dimension where an analytic approach is lacking and demonstrate the lack of dependence of the CFR scaling on the details of the diffusion. Finally we extend the discussion to higher order aggregation processes (Sec.~\ref{sec-nary}). We close with a brief summary of the results.

\section{\label{sec-model} Model Definition and Heuristic CFR}

Consider a $d$-dimensional hypercubic lattice occupied by point size 
particles carrying a positive mass. Multiple occupancy of a site is 
allowed. Given a certain configuration, the system evolves in time via 
the following processes. 
\begin{itemize}
\item {\bf Diffusion}: A particle hops with a mass dependent diffusion rate $D(m)$ to a randomly 
chosen nearest neighbour. 
\item{\bf Coagulation}:
Two particles of masses $m_1$ and $m_2$ on the same lattice site coagulate at rate $\lambda(m_1, m_2)$ to form a particle of mass $m_1+m_2$. 
\item{\bf Input}: 
Particles of mass $m_0$ are injected at rate $J/m_0$ uniformly 
and independently in space. 
\end{itemize}
The initial condition is one where the 
lattice is empty. We shall call this model the mass model (MM).

We will restrict ourselves to the case where the reaction rate 
$\lambda(m_1,m_2)$ is a homogeneous function of its arguments, 
i.e., 
\begin{equation} 
\lambda(\Lambda m_1, \Lambda m_2) = 
\Lambda^{\beta} \lambda(m_1, m_2), 
\end{equation} 
where $\beta$ is the homogeneity exponent. The diffusion constant, $D(m)$, 
will be assumed to have the property 
\begin{equation} 
\frac{D(m)}{D(m_0)} = \left( 
\frac{m}{m_0} \right)^{\kappa}. 
\end{equation} 
Thus, in addition to the different rates, the model has $2$ parameters: the 
homogeneity exponent $\beta$ and the diffusion exponent $\kappa$. In the large time limit, as described in the introduction, this model tends to a statistically stationary state characterised by a constant average flux of mass from small clusters to large ones.

In \cite{CRZ2007} we presented quite a general argument to determine the scaling of the flux-carrying correlation function for a broad class of non-equilibrium systems which reach a constant flux stationary state. In this article, in the interest of clarity, we will briefly review the argument heuristically for the specific case of particle aggregation.

Schematically (we shall write down an accurate expression in Sec. \ref{sec-cfr}), the transfer of mass between coalescing clusters is described by an equation of the form:
\begin{eqnarray}
\label{eq-heuristicHopf}
\nonumber &&\pd{ }{t} \langle m N_m(t) \rangle =\pd{J_m}{m}\\
&\sim& \int  d m_1 d m_2\, m\,\lambda(m_1, m_2)\, C(m_1,m_2)\, \delta_{0;1,2}
,
\end{eqnarray}
where $\delta_{0;1,2}$ is shorthand notation for $\delta(m-m_1-m_2)$.
The right hand side defines the mass flux, $J_m$, in the space of cluster sizes.
$C(m_1,m_2)$ is proportional to the probability of having two clusters with masses $m_1$ and $m_2$ meet at the same point in space. This is the flux-carrying correlation function since it mediates the transfer of mass in the system. Note that the flux-carrying correlation function is not an esoteric object. It has a clear and intuitive physical meaning

In the statistically stationary state, $\pd{N_m(t)}{t}=0$ so that $J_m$ is a constant, independent of $m$. Simply counting powers of $m$ would then lead us to expect that
\begin{equation}
\label{eq-cfr}
C(m_1, m_2) \sim m^{-\beta-3}.
\end{equation}
This heuristic scaling argument is the CFR at the most basic level: mass conservation fixes the scaling of the flux-carrying correlation. The remainder of the paper will be devoted to making this heuristic argument precise and identifying its limitations.

\section{\label{sec-cfr} Improving on the Heuristic CFR}

In this section we arrive at Eq.(\ref{eq-cfr}) more carefully.
Starting from the lattice model, it is relatively straightforward to write down the
evolution equation for the different correlation functions. A full exposition  can be found in \cite{CRZ2006}. Skipping the details, we write directly the equation for $\langle N(m, \xv, t) \rangle$, the average number of particles of mass
$m$ at position $\xv$ at time $t$:
\begin{eqnarray}
\label{mmhopf}& &  \left( \pd{}{t} - D(m)\,\nabla^2\right) \langle N(m)\rangle = \frac{J}{m_0} \delta(m-m_0)\\
& &+\int_0^\infty d m_1 d m_2\, \lambda(m_1, m_2)\, C(m_1,m_2)\, \delta_{0;1,2} \nonumber \\
& &-\int_0^\infty d m_1 d m_2\, \lambda(m,m_1)\, C(m, m_1)\, \delta_{2;01} \nonumber\\
& &-\int_0^\infty d m_1 d m_2\, \lambda(m_2,m)\, C(m_2, m)\, \delta_{1;20}. \nonumber
\end{eqnarray}
For simplicity, we suppress $\vec{x}$ and $t$ dependences and adopt the 
reduced notation for the $\delta-$ functions defined after Eq.(\ref{eq-heuristicHopf}). 
$C(m_1,m_2)$, the flux-carrying correlation function, is defined:
\begin{eqnarray} 
\label{eq-fluxCarryingCorrelationFunction} C(m_{1}, m_{2}) &=&\langle
N(m_{1},\xv,t)N(m_{2},\xv,t)\rangle \\
 & &- \frac{1}{\Delta
x^{d}}\delta(m_{1}-m_{2})\langle N(m_{1},\xv,t) \rangle, \nonumber
\end{eqnarray}
$\Delta x$ being the lattice spacing. Let us explain the terms in Eq~(\ref{mmhopf}) one by one.

The $\nabla^2$ term accounts for particle diffusion which may be mass dependent. For spatially homogeneous statistics, this term is zero.  The first term on the right hand side accounts for influx of particles of mass $m_0$.  The remaining terms account for aggregation processes. To explain the meaning of $C(m_1,m_2)$, we first consider how it
relates to the mean-field Smoluchowski equation. Mean-field theory requires two assumptions. Firstly, correlations are absent so we may write $\langle N(m_{1},\xv,t)N(m_{2},\xv,t)\rangle $ as a simple product of densities, $\langle N(m_1) \rangle \langle N(m_2)\rangle$. Secondly, densities are high so we may neglect the $\langle N(m_1) \rangle$ term relative to $\langle N(m_1) \rangle \langle N(m_2)\rangle$. In the diffusion--limited regime, $C(m_1,m_2)$ has an important probabilistic interpretation.  Writing the averaging process explicitly:
\begin{eqnarray*}
C(m_{1}, m_{2}) &=& \hspace{-0.25cm} \sum_{N_1,N_2=1}^\infty \hspace{-0.25cm}{\mathbb P}(N(m_{1},\xv) = N_1, N(m_{2},\xv) = N_2)\\
&& \times \left( N_1 N_2 - \delta_{m_1, m_2} N_1 \right).
\end{eqnarray*}
This is the average number of pairs of particles with masses $m_1$ and $m_2$ on a site, with the delta function accounting for double counting of particles of equal mass. In the low density (diffusion--limited) regime,
\begin{equation}
C(m_{1}, m_{2}) \approx {\mathbb P}(N(m_{1},\xv) = 1, N(m_{2},\xv) = 1),
\end{equation}
the probability that two particles of masses $m_1$ and $m_2$ meet at a site. Thus the flux-carrying correlation function is not an esoteric object and has a very natural physical meaning. Having understood the meaning of $C(m_{1}, m_{2})$, the second term on the right-hand side of Eq.~(\ref{mmhopf}) accounts for the creation of particles of mass $m$ at $\xv$  through aggregation of 2 particles at $\xv$. The third  and fourth terms account for the decrease of  $N(m,\xv,t)$ through aggregation with other particles.
These latter two terms are identical under relabeling $(m_1, m_2) \to (m_2,m_1)$ and
are usually written as a single term. We write them this way for reasons which will
become obvious below.
 
To simplify the equations,
we introduce $I(m_1,m_2;m)$ defined as:
\begin{equation}
\label{eq-defI}
I(m_1,m_2;m) = \lambda(m_1,m_2)\, C(m_1, m_2)\, \delta_{0;1,2} 
\end{equation}

As already mentioned,  in Eq.~(\ref{mmhopf}) the diffusion term drops out by spatial homogeneity. Then, for $m>m_0$ we can write Eq.~(\ref{mmhopf}) as
\bea
\label{eq-hopf2}
\lefteqn{\pd{\langle  N(m)\rangle }{t} =
\int_0^\infty dm_1 d m_2 \bigg[ I(m_1, m_2;m)} 
\nonumber \\
&&- I(m_2, m ; m_1)  - I(m, m_1 ; m_2 \bigg]
\label{mmhopf1}
\eea
In the steady state, we set the left hand side to zero. To solve this 
equation, we
need to balance out the plus and the minus terms on the right hand. As written, 
it is difficult to see what the
solution is because each term comes with a different delta function.
The balance can be made explicit by the Zakaharov transform (ZT) \cite{ZLF92}. 

Leave the first term as it is. Make the following transformation of the second integral
\begin{eqnarray}
\label{eq-ZT}
m_1 &\rightarrow& \frac{m m_1}{m_2},\\
m_2 &\rightarrow& \frac{m^2}{m_2}.
\end{eqnarray}
The Jacobian of the transformation is $(m/m_2)^{3}$.
Perform the analogous transformation of the third integral (see \cite{CRZ2004}).
Now look for homogeneous solutions, $i.e.$,
\begin{equation}
C(\Lambda m_{1}, \Lambda m_{2}) = \Lambda^h C(m_{1},
m_{2})
\end{equation}
Using this and the homogeneity of $\lambda$, we obtain
\bea
0= \! \int_{0}^{\infty} dm_1 d m_2\, I(m_1, m_2; m)\, (m^y-m_1^y-m_2^y) \label{cfTM}
\eea
where $y=-h-\beta-2$. 

Due to the delta function in Eq(\ref{eq-defI}), I is non-zero only when $m_1 + m_2 =m$.  
If the term in the square bracket is zero when $I$ is non-zero,  
then the equation is satisfied.  Thus, $y=1$ is a solution. This implies
\begin{equation}
h= -\beta-3.
\label{cfrTM}
\end{equation}
It can be easily shown that this is the unique homeogeneous stationary solution of Eq.~(\ref{eq-hopf2}). Introducing rescaled variables, $x_1=m_1/m$ and $x_2=m_2/m$ and
using the assumed homgeneity of $C(m_1,m_2)$, Eq.~(\ref{cfTM}) can be rewritten as
\begin{equation}
0=m^{1+h+\beta+y}\int_{0}^{1} d x_1 d x_2\, I(x_1, x_2; 1)\, (1-x_1^y-x_2^y)
\end{equation}
Due to the delta function in $I(x_1, x_2; 1)$, the integrand is zero unless $x_1+x_2=1$. When $x_1+x_2=1$, the integrand clearly vanishes for $y=1$. To show that this is the only value of $y$ for which the integral is zero, we show that for $y\neq$ the integrand is sign definite on the domain of integration so that the integral is not zero. From the definition, Eq.~(\ref{eq-defI}), $I(x_1, x_2; 1)$ is clearly positive. It remains to consider the function $f(x_1,x_2) = 1-x_1^y-x_2^y$. For $y>1$ the fact that $x_i \in (0,1)$ implies that $x_i^y<1$ so that $x_1^y + x_2^y < x_1 + x_2 =1$. Thus $f(x_1,x_2)>0$ and the integrand is everywhere positive. Likewise, for $y<1$, $x_i \in (0,1)$ implies that $x_i^y>1$ so that $x_1^y + x_2^y > x_1 + x_2 =1$. Thus $f(x_1,x_2)<0$ and the integrand is everywhere negative. Thus, for $y\neq 1$ the integral
does not vanish and the only solution is $y=1$.

One may make a curious observation: the diffusion constant does not play
any role. This is counter to the usual intuition in reaction diffusion systems which holds that diffusion is unimportant for dimensions greater than 
upper critical dimension and all-important for dimensions lower.
Here, we have shown that the 2-point correlation function is
independent of dimension and of the spatial transport mechanism.

It must be pointed out that these manipulations are correct 
provided each of the integrals in the evolution equation are convergent. 
This condition referred as the locality condition has to checked separately. This will be discussed next.

\section{\label{sec-locality} Locality: When is CFR realisable? }
To obtain the formal scaling solution [Eq.~(\ref{cfrTM})] for the evolution
equation (Eq.~(\ref{mmhopf})), some implicit assumptions were made. These
assumptions will be referred to as locality condition, the terminology being
borrowed from wave turbulence.  Unless, these assumptions can be proved or 
checked numerically, the scaling solution should not be expected to hold.  In this section, we explain the locality condition in detail.

For the scaling solution with exponent given by Eq.~(\ref{cfrTM})
to be physically realisable, it must
yield a convergent integrand on the right hand side of Eq.~(\ref{mmhopf}), {\em before}
any changes of integration order are made. Otherwise, divergences cancel leaving a finite contribution.
\begin{figure}
\includegraphics[width=5.5cm]{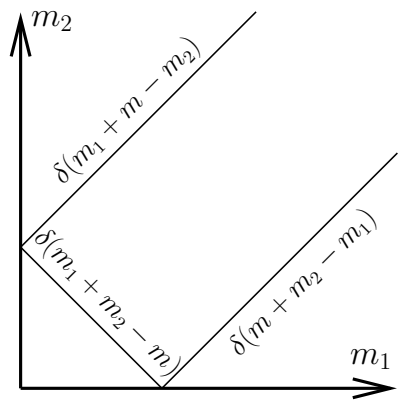}
\caption{\label{fig1} The support of the integrand
of Eq.~(\ref{mmhopf}) in the $m_1, m_2)$ plane.}
\end{figure}

To study this, let us write the two
point function as
\begin{equation}
C(m_1, m_2) = (m_1m_2)^{h/2} \phi\left(\frac{m_1}{m_2}\right),
\label{scalingformTM} 
\end{equation}
thus introducing the dimensionless scaling function $\phi(x)$. 
$\phi(x)$ has the symmetry property
$\phi(x)=\phi(x^{-1})$. To check convergence, it is not enough
to know just the degrees of homogeneity  but rather we require
to know limiting behaviour of various quantities in the integrand.
Suppose
\begin{eqnarray}
\label{eq-scalingFunctionExponents}
\lambda(m_1, m_2) &\sim& m_1^\mu m_2^\nu, \quad \mbox{for $m_2 \gg m_1$},\\
\phi(x) &\sim& x^\sigma, \quad \quad \mbox{for $x \ll 1$}.
\end{eqnarray}
The exponents $\mu$ and $\nu$ are determined by the
model under consideration and must satisfy $\mu+\nu=\beta$.
The behaviour of the scaling function $\phi(x)$ as $x\rightarrow 0$,
as
determined by the exponent $\sigma$, is something which we do not
apriori know. 

The support of the integrand in Eq.~(\ref{mmhopf}) is shown in Fig.~\ref{fig1}. 
We may integrate once and consider the integral as an integral in $m_1$ only.
By scale invariance, we need to check convergence only at the endpoints
of the range of integration. The analysis was done in Ref.~\cite{CRZ2004} in the mean
field limit. Following the analysis of \cite{CRZ2004}, as $m_1\to\infty$, the
behaviour of the integrand is given by
\begin{displaymath}
\lambda(m, m_1) (m m_1)^{h/2} \phi\left(\frac{m}{m_1}\right)  \!  \sim m^\mu
m_1^\nu (m m_1)^{h/2} \!\left(\frac{m}{m_1}\right)^\sigma.
\end{displaymath}
The integral is convergent at infinity if
\begin{equation}
\label{eq-lowerLimit}
-h/2 > \nu+1-\sigma. 
\end{equation}
For the behaviour at $m_1 \rightarrow 0$, 
there is a cancellation of leading order terms :
\begin{eqnarray*}
&&\lambda(m_1,m-m_1)C(m_1, m-m_1)-\lambda(m_1,m)C(m_1,m)\\
&&\sim m_1\frac{\partial}{\partial x}
\left[ \lambda(m_1, x) C(m_1, x)\right]_{x=m} + o(m_1^2),\\
&&\sim m_1 \frac{\partial}{\partial x}\left[ m_1^\mu x^\nu
(m_1x)^{h/2} \phi\left(\frac{m_1}{x}\right)^\sigma\right]_{x=m} +
o(m_1^2).
\end{eqnarray*}
The integral is convergent at $0$ if
\begin{equation}
\label{eq-upperLimit}
-h/2<2+\mu+\sigma. 
\end{equation}
Putting together Eq.~(\ref{eq-lowerLimit}) and Eq.~(\ref{eq-upperLimit}), a convergent collision integral requires that the interval $[\nu+1-\sigma, \mu+2+\sigma]$ should have positive width.  The width of this interval is $2\sigma +\mu-\nu+1$. Thus a convergence requires
\begin{equation}
\sigma > \frac{1}{2}(\nu -\mu -1).\label{loccrit}
\end{equation}
It is easy to show that if this interval exists, the exponent $-h/2$ lies within it assuring the validity of the CFR solution.

At the level of mean field theory, $\sigma=0$ since $C(m_1,m_2)$ is simply proportional
to the product of the 1-point densities. This case was worked out in
detail in Ref.~\cite{CRZ2004} and is consistent with Eq.(\ref{loccrit}).

Thus the rigorous verification of CFR in MM requires
the knowledge of the small-$x$ behaviour of the scaling function.
The latter can be often studied using perturbative methods. For
instance, consider constant kernel MM, $\mu=\nu=0$. If
dimension of the physical space is two, mean field approximation
is applicable, perhaps modulo logarithmic corrections. Within mean 
field
approximation, $\sigma =0$ and criterion Eq.~(\ref{loccrit}) is
satisfied. Hence, CFR holds in two dimensions, meaning in this
case, $C(m_{1}, m_{2}) \sim m^{-3}$, and logarithmic corrections
are absent. Consider now constant kernel MM  in
$d=2-\ep$, where $\ep>0$. The dynamics of the model is governed
now by a fixed point of renormalization group. The order of the
fixed point is $\ep$. Scaling exponents can be now computed using
$\ep$-expansion. As $\sigma (\ep=0)=0$, $\sigma
(\ep)=\sigma_{1}\ep+O(\ep^2)$, where $\sigma_{1}$ is a constant.
Assuming that $\sigma_{1}\neq 0$, one can re-write locality
criterion as
\begin{eqnarray}
\ep \sigma_1>-\frac{1}{2},
\end{eqnarray}
which is  satisfied, if $\ep$ is small enough. In this case
all $\ep$-corrections to $h=-3$ vanish and CFR holds.

This perturbative argument demonstrates that systems which are local
at the mean-field level remain so for some time as one decreases the
physical dimension into the diffusion-limited regime. It does not
tell us much about whether locality holds by the time one reaches
the next physically relevant, integer valued dimension
below the critical dimension since this is presumably beyond the
perturbative regime. In general, this is a difficult problem. The only
case which we are aware of which can be handled analytically is the 
constant kernel ($\beta=0$). It may be shown to be true for $d>2$ where mean field holds \cite{CRZ2004}, in $d=2$ due to a cancellation of
logarithmic corrections \cite{CRZ2006} and in $d<2$ by an exact
solution \cite{RM2000}.

Lacking an analytic approach for other kernels, one must rely on numerical 
simulations to {measure} the value of $\sigma$ for particular systems. 
We perform a systematic numerical investigation of locality in one dimension
for several kernels in Sec. \ref{sec-numerics-locality}.

\section{\label{sec-simulations} Numerical Simulations of CFR}
\subsection{Numerical measurements of CFR exponent}
\begin{figure}
\includegraphics[width=\columnwidth]{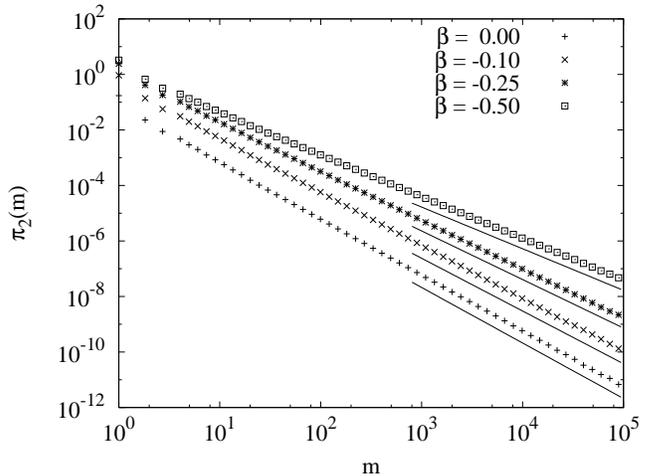}
\caption{\label{fig-CFRNegativeBeta} The variation of $\pi_2(m)$ with $m$ is for
$\beta=-0.5$ (top curve), $-0.25, -0.10, 0$ (bottom curve). The solid
lines correspond to the exponent as predicted by CFR [see Eq.~(\ref{cfrTM})]. Curves have been slightly shifted for
clarity. 
}
\end{figure}

\begin{figure}
\includegraphics[width=\columnwidth]{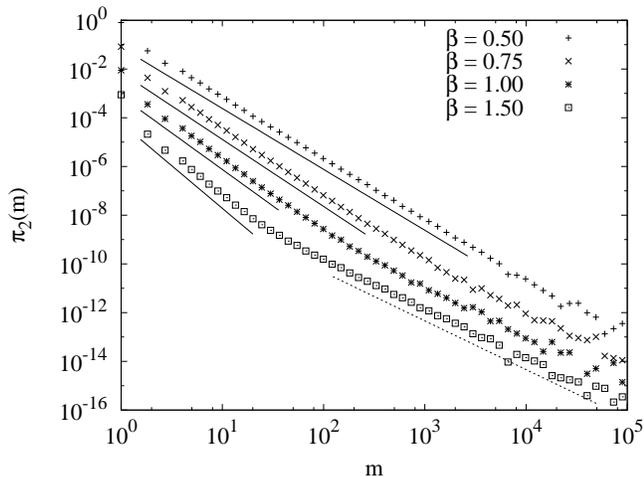}
\caption{\label{fig-CFRPostiveBeta} The variation of $\pi_2(m)$ with $m$ for
$\beta=0.50$ (top curve), $0.75, 1.00, 1.50$ (bottom curve). The solid
lines correspond to the exponent as predicted by CFR [see Eq.~(\ref{cfrTM})].
The dotted line corresponds to exponent for $\beta=0$.
Curves have been slightly shifted for clarity. 
}
\end{figure}

In this section, we present results of numerical
simulations directly measuring the exponent $h$ given in Eq.~(\ref{cfrTM})
It is in one dimension that the effects of fluctuations are the strongest.
Hence, if the mean field scaling predicted by CFR is violated, then it will
be violated in one dimension too. For this reason, all the numerical results that we
show will be monte carlo simulations for one dimensional lattices.

In our simulations we investigated the following representative kernels:
\begin{eqnarray}
\label{eq-additiveKernel}\lambda(m_1,m_2) &=& m_1^\beta + m_2^\beta,\\
\label{eq-muliplicativeKernel}\lambda(m_1,m_2) &=& (m_1 m_2)^{\beta/2},\\
\label{eq-mixedKernel}\lambda(m_1,m_2) &=& \max[m_1, m_2]^\nu \!\min[m_1,m_2]^\mu,
\end{eqnarray}
which we shall refer to as the additive, multiplicative and mixed kernels
respectively. As far as the value of $h$ was concerned, the results were identical for all three kernels. Hence, unless stated otherwise, our figures present results only for one of them, namely the additive kernel, Eq(\ref{eq-additiveKernel}).

What is convenient to measure in simulations is not $\langle
N(m_1) N(m_2) \rangle$, but the quantity,
\begin{equation}
\pi_2(m) = \int_m^\infty dm_1 \langle N(m) N(m_1) \rangle. 
\end{equation}
CFR predicts that $\pi_2(m)$ scales 
as $\pi_2(m) \sim m^{-2-\beta}$. 

In Fig.~\ref{fig-CFRNegativeBeta}, the variation of $\pi_2(m)$ with $m$ is shown for $\beta = 0, -0.10,-0,25,-0,50$. The solid lines are the CFR results. As can be
seen, there is excellent agreement, confirming that CFR holds when $\beta\leq
0$. Fig.~\ref{fig-CFRPostiveBeta}, shows the variation of $\pi_2(m)$ with $m$ for $\beta = 0.50,0.75,1.00, 1.50$. The solid lines are the CFR results. The
CFR exponent is obtained for small and intermediate masses but there is a clear cross-over to another behaviour at large masses. This can be understood as a finite size effect which we shall discuss in section \ref{sec-numericsFiniteSizeEffects}.

At this point it is appropriate to make some comments on the correspondence with mean field theory. For the additive kernel, Eq(\ref{eq-additiveKernel}), at the mean field
level it is believed \cite{VDO1987} that $\beta=1.0$ corresponds to the threshold for instantaneous gelation so that analytical understanding of the solutions of the Smoluchowski equation for $\beta>1$ is very difficult. Notwithstanding the cross-over to another regime at large masses, it is very interesting that the CFR exponent is
observed over some considerable range even for $\beta>1$. To the best of our knowledge,
nothing is known about the behaviour of gelling kernels in the diffusion limited regime or in the presence of a monomer source. In this light, the results of Fig.~\ref{fig-CFRPostiveBeta} pose many interesting questions such as whether there is any remnant in the diffusion--limited regime of the catastrophic singularity which occurs in the mean--field equation at $\beta=1$.

\subsection{\label{sec-numericsFiniteSizeEffects}Finite size effects}

\begin{figure}
\includegraphics[width=\columnwidth]{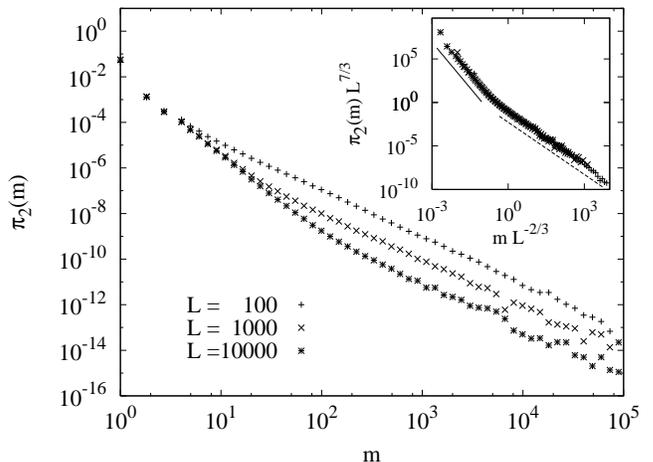}
\caption{\label{fig4} The lattice size dependence of $\pi_2(m,L)$ is shown
for $n=3$ and $\beta=1.5$. The different lattice sizes are
$L=100,1000,10000$. The crossover moves to the right with increasing $L$.
Inset: The curves are scaled as in Eq.~(\ref{finitesize}). The straight lines
correspond to slopes $-7/2$ and $-2$.
}
\end{figure}

Let us consider why the lattice size should affect the CFR scaling. At this
point it is useful to recall two things. Firstly, the recurrence
property of random walks plays a crucial role in determining the statistics of aggregation in the diffusion limited regime. Due to recurrence, heavy particles develop
``zones of exclusion'' around them as they grow, resulting in strong anti-correlations
between heavy particles.  Secondly, recall that the CFR exponent quantifies the decreasing probability of two heavy particles meeting each other. Due to the presence
of zones of exclusion, this probability decreases faster for heavy particles than
the product of one-point densities would suggest.

Although zones of exclusion grow larger as particles
get heavier, in an infinite system there are always enough heavy particles to
maintain the CFR scaling over all mass scales. In a finite system, however, these zones of exclusion become limited by the system size eventually. Once this happens, heavy particles start to meet each other more often than would be expected from CFR since thay can no longer grow their zones of exclusion any larger. Thus a finite size cross-over occurs and results in a shallower scaling as evident from Fig.~\ref{fig-CFRPostiveBeta}. No such crossover occurs for the $\beta<0$ cases shown in Fig.~\ref{fig-CFRNegativeBeta} since for $\beta<0$ large mass clusters become progressively less reactive which acts to counterbalance the growth of zones of exclusion due to recurrence.

The argument above does not explain why finite size effects should lead to a the scaling corresponding to $\beta=0$ indicated in Fig.~\ref{fig-CFRPostiveBeta}. We suggest the following heuristic argument. For a finite system size, $C(m_1,m_2)$ for 'large' masses is contributed to by configurations consisting
of two heavy particles which have been in the system for times  $\gg L^2$, so that 
they are strongly-anticorrelated. Hence these two particles effectively
interact with each other at infinite rate, with effective diffusive jumps of the size
equal to system size.  Hence $C(m_1,m_2)$ behaves as if beta=0 at these masses. Since the mass flux is carried by the meetings of these super-heavy particles, it
is presumably highly intermittent. It is then intuitive that the constant flux argument should fail to describe this regime.

Given that we expect to see CFR scaling for small masses and $\beta=0$ behaviour for large masses, we expect that $\pi_2(m,L)$ should have the form:
\begin{equation}
\pi_2(m,L) = \frac{1}{L^{1+2/\beta}} f \left( \frac{m}{L^{1/\beta}} \right), 
\quad \beta>0,
\label{finitesize}
\end{equation}
where the scaling function $f(x)$ varies as $f(x) \sim x^{-2-\beta}$ when $x
\rightarrow 0$ and $f(x) \sim m^{-2}$ when $x \gg 1$. The crossover mass $m_c$ is given by
$m_c^\beta \sim L$, or $m_c \sim L^{1/\beta}$.

In Fig.~\ref{fig4}, we study the variation of $\pi_2(m,L)$ with $m$ for fixed
$\beta$ and different $L$. The $\beta$ value is chosen to be
$\beta=1.5$. As expected from the preceeding discussion, the crossover point moves to the right with increasing $L$. In
the inset, the data is scaled according to Eq.~(\ref{finitesize}) and excellent
collapse is obtained.  The large and small $x$ behaviour of the scaling function behaves as
predicted.

\subsection{\label{sec-numerics-locality}Numerical validation of locality criterion}

\begin{table}
\begin{ruledtabular}
\begin{tabular}{r|ccccccc} \backslashbox{$\beta$}{$\nu$} 
& -0.250 &-0.125 & 0.000 & 0.125 & 0.250 &0.375 & 0.500 \\ \hline
-0.25 &1.33 & 1.33 & 1.35 & 1.44 & 1.57 & 1.69 & 1.82  \\
 0.00 &1.34 & 1.32 & 1.34 & 1.45 & 1.58 & 1.70 & 1.83  \\
 0.25 &1.32 & 1.31 & 1.34 & 1.46 & 1.58 & 1.70 & 1.83  \\
 0.50 &1.31 & 1.31 & 1.33 & 1.46 & 1.59 & 1.70 & 1.82  
\end{tabular}
\end{ruledtabular}
\caption{\label{table1} The numerical values of $\sigma-h/2$ 
are shown for different $\nu$ and $\beta$. The
kernel used in $\lambda(m_1,m_2) = \max(m_1,m_2)^\nu
\min(m_1,m_2)^\mu$. The errors in the values are $\pm 0.02$.}
\end{table}

As stressed in Sec.~\ref{sec-locality}, aside from a couple of special cases we do not know whether the locality criterion is satisfied in one dimension or not.
We now present numerical measurements of the exponent $\sigma$ in Eq.(\ref{eq-scalingFunctionExponents}) to address this issue. We choose our lattice size sufficiently large to avoid any question  of the finite size effects discussed in the previous section influencing the exponents. We use the mixed kernel, Eq.(\ref{eq-mixedKernel}) so as to be able to vary $\nu$ and $\mu$ independently. What is measured numerically is $\langle N(m_1) N(m_2)
\rangle $ when $m_1$ is kept fixed and $m_2 \gg m_1$. 
Then $\langle N(m_1) N(m_2) \rangle \sim m_1^{h/2+\sigma} m_2^{h/2 -
\sigma}$. In our simulations we keep $m_1$ fixed at $5 m_0$ and take $m_2$
large and measure $\sigma-h/2$. The results of a systematic set of numerical experiments are shown in Table~\ref{table1}.
What one sees is that $\sigma-h/2$ is independent of $\beta$ and dependent
only on $\nu$. The numerics suggest that:
\begin{equation}
\sigma-\frac{h}{2} = \frac{4}{3} + \max[\nu,0].
\end{equation}
If this is true, then $\sigma= -1/6 + (\nu-\mu)/2$ when $\nu>0$,  and
$\sigma= -1/6-\beta/2$ when $\nu<0$. Comparing with the
locality condition in Eq.~(\ref{loccrit}), we see that it is always
satisfied. Based on numerical evidence we therefore conclude that the spatially extended system
is able to adapt itself to variations in the exponents $\mu$ and $\nu$ so that
the locality criterion is always satisfied.

\subsection{Lack of dependence on spatial transport mechanism}

\begin{figure}
\includegraphics[width=\columnwidth]{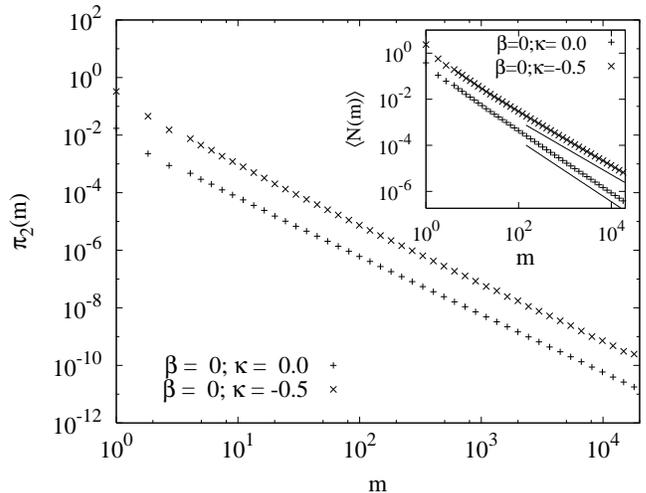}
\caption{\label{fig5} The dependence on diffusion constant of 
of $\pi_2(m,L)$ is shown
for $n=3$ and $\beta=0$. The bottom curve corresponds to $D(m) \sim m^0$
[$\kappa=0$] and
the top curve corresponds to $D(m) \sim m^{-1/2}$ [$\kappa=-1/2$]. 
The curves have same
slope. Inset: The variation of $\langle N(m) \rangle$ with $m$ is shown.
There is a strong dependence on $\kappa$. The solid lines have slope
$(4+\kappa)/3$.
}
\end{figure}

An important prediction of CFR is the lack of dependence on
the diffusion constant. 
In Fig.~\ref{fig5}, we show two sets of data for the same kernel but
different diffusion constants. In one the diffusion constant is independent of
mass. In the other it goes as $D(m) \propto m^{-1/2}$ such that $\kappa =
-1/2$. As can be seen, $\pi_2(m)$ scales exactly the same. However, as shown
in the inset, the one point distribution function $\langle N(m) \rangle $
scales differently with $\kappa$.

\section{\label{sec-nary} Higher order aggregation processes}
Higher order aggregation processes may be considered where coalescence can only
occur when $n-1 > 2$ particles meet at a single site. Although such processes have fewer physical applications than the binary case ($n=3$), they have been suggested as an appropriate model of certain polymeric reactions \cite{JGB90} and have received some attention in the literature \cite{JG89,K94,LEY2003}. From a theoretical perspective, such systems provide an illustrative example of the breakdown of the CFR scaling due to a violation of the locality criterion. For these reasons, we consider the extension of the CFR argument to such systems.

Again, we will restrict ourselves to the case where the reaction rate 
$\lambda(m_1, \ldots, m_{n-1})$ is a homogeneous function of its arguments of degree $\beta$. The Hopf equation corresponding to Eq.(\ref{mmhopf}) is:
\begin{widetext}
\bea
 \left( \pd{}{t} - D \nabla^2\right) N(m) &=& 
 \int_0^\infty    \prod_{i=1}^{n-1} dm_i
\lambda(m_1,\ldots,m_{n-1}) \prod_{i=1}^{n-1} N(m_i)  
\delta\left[ \sum_{i=1}^{n-1} m_i -m \right] \nonumber \\
&&-(n-1) \int_0^\infty   \prod_{i=1}^{n-2} dm_i
\lambda( m_1,\ldots, m_{n-2,m}) N(m) \prod_{i=1}^{n-2} N(m_i)  +
\frac{J}{m_0} \delta(m-m_0),
\label{eq-naryHopf}
\eea
\end{widetext}
The flux-carrying correlation function is the $(n-1)$-point correlation function denoted by
\begin{equation}
C(m_1, \ldots, m_{n-1}) = \langle N(m_1) \ldots N(m_{n-1}) \rangle, \quad m_i
\neq m_j.
\end{equation}
By analogy with Eq.(\ref{eq-defI}), we introduce a quantity $I(m_1,\ldots,m_{n-1};m_{n})$:
\begin{eqnarray}
\lefteqn{I(m_1,\ldots,m_{n-1};m_{n}) =}
\nonumber \\
&& \!\!\! \!\!\lambda(m_1, \ldots, m_{n-1}) 
C(m_1, \ldots, m_{n-1}) 
\delta\!\left[\sum_{i=1}^{n-1} m_i -m_{n}\right]
\end{eqnarray}
On taking average in Eq.~(\ref{eq-naryHopf}), the diffusion term drops out. 
Then, for $m>m_0$ we can write Eq.~(\ref{eq-naryHopf}) as
\begin{eqnarray}
\lefteqn{\pd{\langle  N(m)\rangle }{t} =
\int_0^\infty \prod_{i=1}^{n-1} dm_i \bigg[ I(m_1, \ldots, m_{n-1};m)} 
\nonumber \\
&&-\sum_{j=1}^{n-1} I(m_1, \ldots, m_{j-1},m, m_{j+1}, 
\ldots m_{n-1}; m_j) \bigg]
\label{eq-naryHopf1}
\end{eqnarray}
The Zakharov transformations are:.
\begin{eqnarray}
m_i &\rightarrow& \frac{m m_i}{m_j}, \quad i\neq j,\\
m_j &\rightarrow& \frac{m^2}{m_j},
\end{eqnarray}
one for each of the $n-1$ negative integrals. They have Jacobians $(m/m_j)^{n}$. Looking for homogeneous solutions,
\begin{equation}
C(\Lambda m_{1}, \Lambda m_{2}, \ldots, \Lambda m_{n-1}) = \Lambda^h C(m_{1},
m_{2}, \ldots, m_{n-1})
\end{equation}
and using the homogeneity exponent of $\lambda$, we obtain
\bea
0= \! \int_{0}^{\infty} \prod_{i=1}^{n-1} dm_i I(m_1, \ldots, m_{n-1}; m)
\left[ m^{y} -\sum_{i=1}^{n-1} m_i^{y} \right] 
\label{eq-naryHopf2}
\eea
where $y=-h-\beta-n+1$.  We obtain a stationary solution when $h=-\beta-n$.
The uniqueness argument of Sec.~\ref{sec-cfr} is easily extended to the case
of n-ary interactions.

It is cumbersome to discuss in full generality 
locality for higher values of $n$. Instead, we do a mean field analysis for
$n=4$ (three particles coalesce to form a new particle) for the additive
kernel $\lambda(m_1,m_2,m_3) = m_1^\beta+m_2^\beta+m_3^\beta$. In the
mean field limit $\langle N(m_1)  N(m_2) N(m_3) \rangle \sim (m_1 m_2
m_3)^{h/3}$. When considering the collision integrals as a function of $m_1$
(say when $m_1 \rightarrow \infty$), there is a free integral over $m_2$.
This integral being an integral over a pure power law, will either diverge at
$\infty$ or at $0$. Hence the integrals are no longer finite and the locality
condition will not be satisfied. Physically what happens is that three body
collisions between three large particles are overwhelmed by three body
collisions involving two large particles and one particle of very small mass.
Thus the system behaves effectively as $n=3$. One can get over this problem
by introducing local kernels as discussed below.

We now present some numerical results for  $n=4$. We consider additive kernel with $\beta=0$, $i.e$,
\be
\lambda(m_1,m_2,m_3) = 1.
\label{nonlocal}
\ee
and measure the quantity
\begin{equation}
\pi_3(m) = \int_m^\infty \int_m^\infty dm_1 dm_2 
\langle N(m) N(m_1) N(m_2)  \rangle. 
\end{equation}
which has a constant flux scaling of $\pi_3(m) \sim m^{-2-\beta}$.

\begin{figure}
\includegraphics[width=\columnwidth]{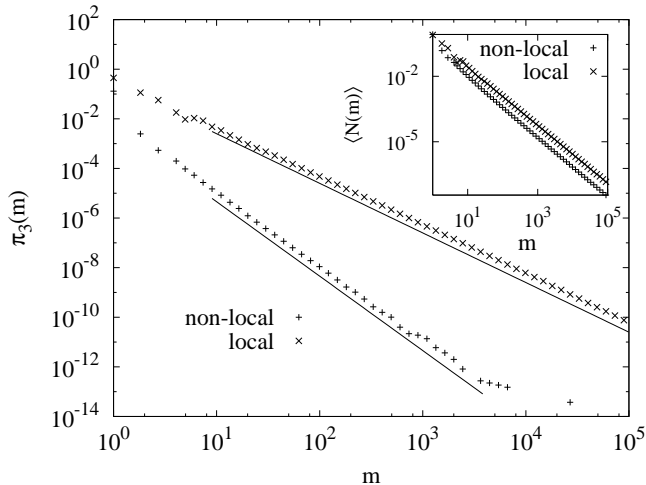}
\caption{\label{fig6} The variation of $\pi_3(m)$ with $m$ is shown for the
non local kernel Eq.~(\ref{nonlocal}) [bottom curve] and the local kernel
Eq.~(\ref{localkernel}) [top curve]. The simulations are for $n=4$. The
solid lines have slope $-3$ and $-2$. Inset: the variation of $\langle N(m)
\rangle$ with $m$ is shown for the local and nonlocal kernels. Their scaling
with $m$ is independent of the kernel for large $m$.
}
\end{figure}

For $\beta=0$, the upper critical dimension is one. Hence, by the argument
above, the locality condition should be violated
and we should get scaling corresponding to $n=3$.
In Fig.~\ref{fig6}, we show the variation of $\pi_3(m)$ with $m$. The bottom
curve correspond to the above kernel. CFR predicts that $\pi_3(m) \sim m^{-2}$
The bottom curve scales as $m^{-3.0}$  corresponding to scaling as predicted
by $\beta=0$ and $n=3$.

To restore CFR, we consider a local kernel of the form
\be
\lambda(m_1,m_2,m_3) = 
g\left( \frac{m_1}{m_2} \right)
g\left( \frac{m_2}{m_3} \right)
g\left( \frac{m_3}{m_1} \right),
\label{localkernel}
\ee
where the dimensionless function is chosen to be
\be
g(x) = \exp\left(x+\frac{1}{x} - 2 \right)
\ee
This local kernel has the effect that it suppresses interactions between
masses that are not of the same magnitude. The results of $\pi_3(m)$ for this
local kernel is presented in the top curve of Fig.~\ref{fig6}. As can be
seen, CFR is now obeyed. The inset of Fig.~\ref{fig6}  shows that for both
the local and non-local kernels $\langle N(m) \rangle$ has the same scaling.
this is again as expected because both for $n=3$ and $n=4$, $\langle N(m)
\rangle \sim m^{-4/3}$ modulo log corrections for $n=4$.

\section{\label{sec-summary} Summary and Conclusions}

To summarise, we have performed an extensive theoretical and numerical study of the applicability and consequences of the CFR argument introduced in \cite{CRZ2007} in the context of cluster-cluster aggregation with a monomer source. We have used a heuristic scaling argument and an exact anaylsis of the appropriate Hopf equation to show that the scaling of the flux-carrying correlation function in the stationary state is fixed by the fact that the elementary coalescence interactions conserve mass. In the case of cluster-cluster aggregation, the flux carrying correlation function is proportional to the probability of $n-1$ clusters coming together at the same point in space. It is thus not an esoteric object but is of direct physical significance. 

The CFR scaling exponent is identical to that given by mean-field theory. It is thus independent of the physical dimension and independent of the details of the spatial transport mechanism. This latter fact we have demonstrated clearly with some numerical simulations of aggregation with mass-dependent diffusion rates. The importance and non-triviality of the result lies in the fact that the flux-carrying correlation function exhibits the mean field scaling even in the diffusion limited regime where mean field theory fails to give correct answers for other correlation functions, in particular for the density. This runs counter to the usual intuition in interacting particle systems where it is canonical that statistics are dominated by diffusive fluctuations in low dimensions where mean field theory breaks down. We do not consider our result to be at odds with this canon. It is indeed the case that most statistical quantities measured in the diffusion limited regime will be fluctuation dominated. What we have shown is that there is a particular special correlation function which does not feel these fluctuations at all.

The usefulness of this result has already been demonstrated in our earlier work \cite{CRZ2005,CRZ2006} on constant kernel aggregation in low dimensions where it allowed us, taken together with a known exact result for the density, to prove multiscaling for the statistics of constant kernel aggregation in one dimension. Given the very direct physical meaning of the flux-carrying correlation function in the aggregation context, it seems likely that other applications will arise in concrete problems. At the very least, one can envisage using the result as a benchmark for numerical simulations of more complicated aggregation problems, much as the $4/5$-Law is used in validating numerical simulations of turbulence.

As we have stated in our earlier paper, the CFR is not a theorem. It requires that a criterion which we refer to as ``locality'' should hold. To reiterate, by ``locality'', we do not mean that only clusters of equal masses are allowed to coalesce (although such a restriction would certainly ensure that our criterion is satisfied). Rather we mean a  much weaker requirement that the mass integrals describing the flux should not be dominated by their upper or lower limits. In general, locality is not testable a-priori. We have therefore devoted a considerable amount of effort in this article to studying the locality criterion in the context of cluster-cluster aggregation. From the theoretical perspective, we showed that if scaling exponents describing a system satisfy locality at the mean-field level (something which {\em can} be checked apriori) then there is a perturbative neighbourhood of models below the critical dimension for which locality holds. We then showed numerically that it is satisfied for a range of kernels in one dimension but breaks down for kernels for which one would expect long-range (in mass space) interactions to become dominant. We provided an instructive illustrative example of how the breakdown in locality may violate CFR using a model kernel where the long range interactions may be tuned.

It is rare that a generic nonequilibrium system will be solvable as the model
discussed in the paper. It could be that the distinction between driving and
dissipation scales get fuzzy \cite{KAD1999}, or it could be that identifying
the conserved quantity is a problem. In a recent paper \cite{CRZ2007B}, we studied a 
model wherein the dissipation scale in not very well defined, and conjectured
a CFR for such a model, even though it would not be expected apriori. The consequences of this conjecture was verified numerically. It would be of interest to clarify these observations theoretically so that the results of the present article might be extended to an even wider class of models.


\end{document}